\documentstyle[preprint,aps]{revtex}
\begin{document}
\title{Commensurate and modulated magnetic phases in orthorhombic $A$C$_{60}$}
\author{E.J. Mele$^1$, G. V. Krishna$^1$, and S.C. Erwin$^2$}
\address{$^1$Department of Physics and Laboratory for Research on the
Structure
of Matter,\\University of Pennsylvania, Philadelphia, Pennsylvania 19104\\
$^2$Complex Systems Theory Branch, Naval Research Laboratory,
Washington DC 20375\\}

\date{\today}
\maketitle
\begin{abstract}
Competing magnetically ordered structures in polymerized
orthorhombic $A$C$_{60}$ are studied.  A mean-field theory for the
equilibrium phases is developed using an Ising model and a classical
Heisenberg model to describe the competition between inter- and
intra-chain magnetic order in the solid.  In the Ising model, the limiting
commensurate one-dimensional and three-dimensional phases are
separated by a commensurate three-sublattice state and by two sectors
containing higher-order commensurate phases. For the Heisenberg model
the quasi-1D phase is never the equilibrium  state; instead
the 3D commensurate phases exhibits a transition to  a continuum of
coplanar spiral magnetic phases.

\end{abstract}


Recent experiments have revealed that $A$C$_{60}$ (where $A$=K, Rb)
undergoes a first-order structural phase transition from a
high-temperature face centered cubic (fcc) structure to a
low-temperature orthorhombic phase \cite{chauvet}.  The structure of
this new low-temperature phase is quite interesting since the lattice
parameter along the polymer direction ($c$ axis) is only 9.1 \AA. This
distance implies a separation between nearest-neighbor molecules along
this axis as small as $\sim$2 \AA, and has led to the suggestion
that the fullerene molecules in this phase polymerize to form an
ordered crystalline array of quasi-one-dimensional (quasi-1D) chains,
resembling a lattice of aligned ``necklaces" of fullerene molecules.
Moreover, below 50 K this orthorhombic phase undergoes a second
electronic phase transition, which Chauvet {\it et al.} argue arises
from a spin-density-wave instability within the quasi-1D chains
\cite{chauvet}.

More detailed electronic structure studies do not support this
interpretation \cite{ekm}.  The conduction band calculated for the
orthorhombic phase shows substantial dispersion for the doped carriers
both parallel and perpendicular to the polymer direction.
Furthermore, by treating the effect of a short-range repulsive
potential on this paramagnetic state within the random-phase
approximation (RPA), one finds that the metallic phase of this
structure favors magnetic fluctuations to a three-dimensionally
ordered two-sublattice state, with spin polarizations alternating on
the corner and body-centered sites of the pseudo-body-centered
orthorhombic cell \cite{ekm}.  Indeed, this spin structure is stable
for quite modest values of the repulsion strength, and is likely
associated with the 50 K transition observed experimentally.

However, it is plausible that a dilation of the interchain galleries,
such as might be achieved by alloying the alkali donors with larger
spacer molecules, could lead to the one dimensional situation
envisioned by Chauvet {\it et al.}  If so, the transition of the
magnetic structure from the three-dimensional ordered phase to the
quasi-1D phase is of some importance, and is the subject of this
paper.  We
study the transition from three-dimensional to one-dimensional
magnetic structures using both an Ising model and a classical
Heisenberg model to describe the spin dynamics.
  Interestingly, as we show below, the passage from the
three-dimensional ordered structure to the quasi-1D structure is not
an easy one.  In fact, at $T=0$ the quasi-1D modulated magnetic phase
appears as an equilibrium phase in the Ising model but not the Heisenberg
model for this system.
In the former case, the ordered quasi-1D magnetic structure is
separated from the three-dimensional structure by an
assortment of intervening modulated magnetic phases, and in the latter case
it is preempted by a continuum of spiral magnetic phases.

The structure of the pseudo-body-centered orthorhombic (pseudo-bco)
phase is sketched in Fig.~\ref{structure}(a).  The measured lattice
parameters (for RbC$_{60}$) are $a$=14.23, $b$=10.11, $c$=9.14 \AA
\cite{stephens}.  There are two candidate orientational structures for
the fullerene molecules on this Bravais lattice; the measured x-ray
diffraction intensities at present are unable to distinguish between a
model with fixed orientations (space group Pmnn) and a model with
binary orientational disorder (space group Immm) \cite{stephens}.  For
simplicity, we assume here a true bco Bravais lattice with a single
orientation \cite{ekm}.  The Brillouin zone for this bco cell is
sketched in Fig.~\ref{structure}(b).  In our model, each Bravais
lattice site of this structure is presumed to localize a single
electron, with the coupling between the sites described by the
Hamiltonian
\begin{equation} H = \frac{1}{2} \sum_{i,j} J_{ij}
\left(S^z_i S^z_j + \frac{\lambda} {2} (S^+_i S^-_j + S^-_i
S^+_j)\right), \label{ham}
\end{equation}
where $S_i^{\pm}=S_i^x \pm iS_i^y$.  We consider the limiting cases of
the Ising model ($\lambda = 0$) and the Heisenberg model ($\lambda =
1$), with the spins treated classically.

The sum in Eq.~(\ref{ham}) runs over the first ten sites $j$
surrounding any lattice site $i$.  This sum thus includes all the
interactions $J_1$ between a reference site and its eight neighbors
along the body diagonal of the bco cell, and the interaction $J_2$ to
the two nearest neighbors along the polymer direction.  Note that this
spin Hamiltonian can be frustrated, since for the choices $J_1 > 0$
and $J_2 > 0$ antiferromagnetic correlations along the body diagonal
must compete with antiferromagnetic correlations along the polymer
chain.  For the actual RbC$_{60}$ structure, our calculations within
the RPA show that the system is in fact not frustrated \cite{ekm}, and
that both of the effective interactions favor the two-sublattice
structure discussed in Ref.~\cite{ekm}, that is, we have a situation
with $J_1 > 0$ and $J_2 < 0$.  More distant interactions within each
of the chains also favor ferromagnetic intrachain order.  However, one
can imagine that that by reducing the size of the interchain hopping
amplitudes one ultimately arrives in the regime with $J_2 >> J_1 > 0$,
which would then favor the quasi-1D spin configuration originally
proposed by Chauvet {\it et al} \cite{chauvet}.  In the following we
label these two limiting commensurate reference states $C_{3D}$ (for
the three-dimensional state) and $C_{1D}$ (for the one-dimensional
state). Schematic representations of these spin configurations are
shown in Fig.~\ref{spinconfigs}.

The general structure of the phase diagram for these colinear states,
shown in Fig.~\ref{phasediagrams}(a), can be understood by comparing the
free energies of the competing colinear states, $C_{1D}$ and $C_{3D}$,
as we vary the interaction ratio $J_2 / J_1$.  Within the Ising model,
mean field theory yields a multicritical line at $J_2 / J_1 = 2$,
which separates the $C_{3D}$ phase from the $C_{1D}$ phase.
Each of these phases is in turn separated from the high-temperature
disordered phase by a second-order phase boundary, denoted by solid
lines.  There are an infinite number of coexisting phases along
vertical dashed line, since the line locates the critical interaction
ratio where the creation energy vanishes for a domain wall which slips
the stacking sequence of the laterally ferromagnetically ordered
layers by one lattice spacing along the polymer axis.

The situation is more interesting if one allows for the possibility
for a longer range modulation of the spin texture.  Consider first the
Ising limit, given by $\lambda = 0$.  The mean-field inverse
susceptibility in the paramagnetic phase of the model has the form
$\chi^{-1}(q,T) = S^2_o J(q) + kT$, where $S^2_o = 1/4$ and $J(q) =
\sum_j J_{ij} \exp(i q \cdot R_{ij})$.  For a given value of
$J_2/J_1$, we then seek the critical wave vector $q_c$ which minimizes
$J(q)$.  Examination of the evolution of $q_c$ reveals that this
system exhibits four limiting instabilities from the paramagnetic
phase: (a) For $J_2/J_1 < 1$, the dominant fluctuations are at the $X$
point of the bco zone, $(2\pi/a,0,0)$; below the critical temperature,
this leads to the $C_{3D}$ spin structure. (b) For $1< J_2/J_1<2$, the
ordering wave vector evolves continuously from the $X$ point towards
the $G$ point, $(2\pi/a,0, 2 \pi/3c)$. (c) For $J_2/J_1 = 2$, the
dominant fluctuation is obtained exactly at the $G$ point.  This leads
to a commensurate three-sublattice antiferromagnetically ordered
phase, $C_{3s}$, which will be briefly discussed below.  (d) For
$2<J_2/J_1$, the modulated structure is again incommensurate, with the
ordering wave vector evolving between the $G$ point and the point, ($2
\pi/a,0,\pi/c)$. The one-dimensional limit of this problem corresponds
to a condensation at any wavevector on the plane $q_z = \pm \pi/c$.

For $0<T<T_c$ we can investigate the properties of the low-temperature
ordered phases by solving self-consistently the linear stability
relations $\sigma_i = \sigma_o \tanh (\beta h_i)$, where the field is
$h_i = - \sum_j J_{ij} \langle \sigma_j \rangle$. The resulting phase
diagram is shown in Fig.~\ref{phasediagrams}(b).  As one expects in
the frustrated Ising model, the degeneracy along the original
multicritical line is broken by the entropic contributions to the free
energy \cite{selke}, and at finite temperature one finds that a $T=0$
multicritical point branches into two sectors containing modulated
incommensurate (actually, higher order commensurate) phases, $I$ and
$I'$, and a locked commensurate three-sublattice phase, $C_{3s}$.
This intermediate commensurate phase is a modulated phase with the
amplitude $\sigma_z (z) = \sigma_o \sin( 2 \pi z / 3c )$. It thus
consists of lateral planes of fullerene molecules ferromagnetically
aligned within a layer, with antiferromagnetic ordering across
isolated pairs of neighboring layers, leaving the third layer in the
cell unpolarized; this is shown schematically in
Fig.~\ref{spinconfigs}.  The regions $I$ and $I'$ are themselves
further partitioned into a denser set of higher order commensurate
phases.

We now remove the scalar constraint on the spin degree of freedom, and
treat the Heisenberg vector-spin model of Eq.~(\ref{ham}) with
$\lambda = 1$.  The orientation of the spin on the $i$-th lattice site
can be specified by an angle $\Omega _i$ locating a point on the
surface of the unit sphere.  Within the mean field theory, expanding
around the disordered phase, we consider a normalized trial density
matrix of the form $\rho (\Omega) = (1/4\pi) \left( 1 + \sum_{\alpha}
p_{\alpha} K_{\alpha} (\Omega) \right)$ with $\alpha = x,y,z$
labeling the three real normalized vector harmonics $(K_x , K_y, K_z)
= \sqrt{3/4\pi}\, (\sin \theta \cos\phi, \sin\theta \sin\phi,
\cos\theta)$.  The thermal average of the $\alpha$-th component is
thus $\langle S_{\alpha} \rangle$ = $\sqrt{1/12\pi}\, S_o p_{\alpha}$.
The mean-field free energy per site in this model has the form
\begin{eqnarray}
 f & = & \frac{1}{2N} \sum_{ij,\alpha} J_{ij,  \alpha} \langle S_{i \alpha}
\rangle  \langle S_{j\alpha}  \rangle \nonumber \\
&& + \frac {kT} {N} \sum_{i, \alpha} \int d \Omega_i \,\rho (\Omega_i) \log
\rho (\Omega_i).
\label{freeen}
\end{eqnarray}
To study the equilibrium magnetic states of the system, we minimize
this trial free energy with respect to the coefficients $p_{\alpha}$.
For $T<<T_c$ the expansion in $K_\alpha$ is inadequate, and we replace
it by an expansion in a set of gaussians uniformly distributed over
the surface of the unit sphere.

To quadratic order in the $p_\alpha$, the free energy in this vector
model takes the form:
\begin{equation}
f =\frac {1}{8 \pi} \sum_{q,\alpha} p_{\alpha} (q) p_{\alpha} (-q)
\left(
\frac {  S_o^2 J(q)}{3} +  kT \right).
\label{vectorf}
\end{equation}
The susceptibility thus diverges on the same locus as found above for
the Ising model, and leads to the same high-temperature phase boundary
as shown in Fig.~\ref{phasediagrams}(b).  The vector character of the
spin qualitatively changes the low-temperature behavior of the model,
however.  At $T=0$ the commensurate $C_{3D}$ phase remains stable only
up to the critical point $J_2/J_1$ = 1.  This can be seen most
directly by studying the long wavelength spin fluctuations around the
ordered $C_{3D}$ state. To do this we adopt a semiclassical model, in
which the equation of motion for a spin precessing on site $i$ is:
\begin{equation}
\frac {d {\bf S}_{i}}{dt} =  {\bf S}_{i} \times  {\bf h}_{e\!f\!f,i}
\label{dynamics}
\end{equation}
where $h_{e\!f\!f, i \alpha} = - \langle \partial H /\partial
S_{i \alpha}\rangle$, and the brackets denote a thermodynamic
average.  Eq.~(\ref{dynamics}) is then solved by linearizing the spin,
${\bf S} \approx \langle {\bf S} \rangle + {\bf \sigma} (t) $. The
dependence of the equation of motion on the geometric structure of the
parent spin state is expressed explicitly in Eq.~(\ref{dynamics}).
Expanding around the ordered $C_{3D}$ phase one finds that the
competition between $J_1$ and $J_2$ renormalizes the spin-wave
velocity.  For $q \rightarrow 0$ the spin waves have the dispersion
$\omega = 8 (J_1 ( J_1 - J_2))^{\frac{1}{2}} q_z c$ , so that at
$J_2/J_1 = 1$ the system is only marginally stable with respect to
transverse fluctuations of the magnetization that are modulated along
the $c$ direction.

For larger values of $J_2/J_1$, the system spontaneously develops a
static transverse undulation of the ordered moment. This yields a
helical phase modulated at a wave vector $q$ that can be tuned
continuously by varying the interaction ratio.  Indeed, for
$J_2/J_1>1$ the equilibrium phase of the model is a coplanar spiral
texture.  This is seen by studying the zero-temperature
internal energy per site, $u$.  We assume that each spin is tipped
with respect to the $c$ axis by a polar angle $\theta$, and that the
transverse component of the magnetization advances by a uniform pitch
$\phi$ between neighboring planes normal to the polymer direction.
The energy per site then is then
\begin{equation}
u =   J_2 \cos^2\theta + J_2 \sin^2\theta \cos 2\phi + 4 J_1 \sin^2
\theta \cos\phi.
\label{energy}
\end{equation}
We see that this interaction energy is always minimized by a texture
with $\theta = \pi/2$ and $\phi = \cos^{-1}(-J_1/J_2)$.  The
dependence of $\phi$ on $J_2/J_1$ is shown in Fig.~\ref{pitch}. One
sees that for $J_2/J_1 \rightarrow 1^+$ the spiral phase smoothly
connects to the ordered $C_{3D}$ phase (this corresponds to the choice
$\theta = \pi/2$ and $\phi = \pi$).  A second critical point occurs at
$J_2/J_1=2$.  Here $\phi$ = $2 \pi / 3$ which describes the
three-sublattice spiral phase. In fact,
the geometric structure of the coplanar three-sublattice spin texture can
already be deduced by a careful inspection of the sixth-order
invariants in the Landau expansion of the free energy for a vector
order parameter near the critical point connecting the
three-sublattice phase to the paramagnetic phase.

Remarkably, the energy of Eq.~(\ref{energy}) requires that for any
finite $J_2/J_1$ (that is, for any nonzero $J_1$) the ground state
will always be a helical structure, albeit with a pitch which diverges
as $J_2 / J_1 \rightarrow \infty$.  This is confirmed by studying the
spin-wave expansion for small fluctuations around a putatively ordered
phase $C_{1D}$ at $T=0$ in the regime $J_2/J_1>2$. Here, global
rotational invariance requires that the $q = 0$ mode occurs at precisely
$\omega = 0$, but for $\it{ nonzero }$ $q$ near the zone center, we find
that for all coupling strengths the transverse fluctuations around
this phase are unstable.  The width of this instability in momentum
space collapses as $J_2/J_1$ diverges.  Thus the commensurate
structure $C_{1D}$ is intrinsically unstable at $T=0$ for all finite
coupling ratios.  In Fig.~\ref{pitch} one sees that the pitch asymptotically
approaches the value $\pi / 2$ , which is its value for the $C_{1D}$
phase, only in the limit of large $J_2/J_1$.

The phase diagram for the classical vector-spin model in the bco
structure is shown in Fig.~\ref{phasediagrams}(c).  The commensurate
$C_{3D}$ phase is stable for $J_2/J_1<1$, and the hatched regions to
the right describe the continuum of coplanar spiral phases by which
the $C_{3D}$ phase winds out, asymptotically approaching the $C_{1D}$
phase.

The energy density of Eq.~(\ref{energy}) requires that the helical
phases are all coplanar spiral textures; thus nowhere in the phase diagram do
we obtain a phase in which the spin texture orders in three dimensions.
However, this can be the case if we treat the effects of a
crystal-field anisotropy.  Orthorhombic symmetry allows an on-site
potential of the form $a S_x ^4 + b S_y^4 +c S_z^4$.  Competition between
this on-site potential and the planar spiral textures favored by the
quadratic interactions can then ``tip" the spins out of the plane,
yielding a stable three dimensionally ordered conical spin configuration.
Of course, in the presence of an external magnetic field,  a weak
external field acts as a director which orients the plane of  the spiral
modulation normal to the applied field, with the spins
canted uniformly along the external field direction.

The analysis given above has been carried out for a model in which the
spins are presumed to be localized on each lattice site of the
structure.  However, it is likely that the doped carriers of the
polymerized solid are better described by the itinerant limit of this
theory, along the lines of the theory discussed in Ref.~\cite{ekm}.
In fact, we note the experimental transition temperature to a
magnetically ordered phase occurs at 50 K, while the Fermi energy for
the doped carries is of order 3000 K, so that quantum fluctuations
around the ordered classical phase can be quite significant for this
system.  Furthermore, we know that the intermolecular hopping
amplitudes in the crystal are very sensitive to the interchain
separation, and are somewhat sensitive the orientational state of the
fullerene molecules.  It would therefore be useful to carry out this
analysis perturbing from the paramagnetic ``metallic" phase of the
doped system (containing both charge and spin fluctuations), as a
function of varying the interchain separations.  The structure of the
resulting phase diagram as well as the character of the low lying
quasiparticle excitations of the condensed phase in the presence of
these ordered spiral spin textures would be particularly interesting
to study.  Even within the classical model, a distribution of
$J_2/J_1$ ratios, such as might be introduced by orientational
disorder on the fullerene sites, can be expected to couple strongly to
the phase degree of freedom of the spiral phases and to lead to
disordered phases with nontrivial magnetic properties which have yet
to be studied.

This work was supported in part by the Laboratory for Research
on the Structure of Matter (University of Pennsylvania), by the
NSF under the MRL program (Grant 92 20668) and by the DOE (Grant 91ER
45118).  Computations were carried out in part at the Cornell Theory
Center, which receives major funding from NSF and New York State.

%
%

%
%

\begin{figure}
\caption {(a) A portion of the pseudo-body-centered orthorhombic
lattice for the polymerized phase. (b) The bco Brillouin
zone and high-symmetry points in the $q_x - q_z$ plane. \label {structure}}
\end{figure}

\begin{figure}
\caption{Spin structures on a cluster from the bco lattice.  The
axes of the polymer chains are indicated by the light
lines.\label{spinconfigs}}
\end{figure}

\begin{figure}
\caption {Phase diagrams for the frustrated spin model
on the bco lattice: (a) phases considering only
colinear structures; (b) phase diagram for the Ising model; (c) phase
diagram for the classical Heisenberg model. \label{phasediagrams}}
\end{figure}

\begin{figure}
\caption {Evolution of the pitch of the coplanar spiral phase as a
function of the interaction ratio. \label{pitch}}
\end{figure}

\end{document}